# Spin-multiplexed terahertz image edge detection based on extrinsic chirality in all-dielectric metasurface

YUXIN TANG,[1,2] SHICHAO ZHAO,[1,2] LI LUO, [1,2] TINGTING TANG, [1,2] HANG XU,[3] JIE LI [1,2,*]

[1] Sichuan Province Key Laboratory of Optoelectronic Sensor Devices and Systems, College of Optoelectronic Engineering (Chengdu IC Valley Industrial College), Chengdu University of Information Technology, Chengdu 610225, China.
[2] Sichuan Meteorological Optoelectronic Sensor Technology and Application Engineering Research Center, Chengdu University of Information Technology. Chengdu 610225, China.
[3]School of Precision Instruments and Opto-Electronics Engineering, Tianjin University, Tianjin 300072, China.
*Corresponding author: li_jie_d@tju.edu.cn.

Extrinsically chiral metasurfaces can exhibit pronounced chiral responses in planar structures without intrinsic chirality, making them attractive for spin-selective light manipulation. Here, we propose a transmissive all-dielectric terahertz metasurface for image edge detection. Owing to the extrinsic chiral response, the metasurface enables edge-enhanced imaging under right-handed circularly polarized incidence near 0.69 THz, while preserving bright-field imaging under left-handed circularly polarized incidence. Spatial-frequency analysis shows that the effective numerical aperture of the device is approximately 0.4, indicating the terahertz image edge detection can be achieved over a broad spatial-frequency range. This work provides a new design route toward transmissive terahertz all-optical image processing with spin selectivity and a large numerical aperture.

Image processing is a key step in information acquisition and intelligent analysis, with broad applications in biomedical imaging, machine vision, and target recognition [1, 2]. Conventional digital image processing usually relies on optoelectronic conversion, digital sampling, and electronic computation, and therefore suffers from latency, power consumption, and system complexity in high-speed real-time applications [1–3]. In contrast, all-optical image processing exploits light propagation and spatial-frequency modulation to perform differentiation, integration, and filtering directly in the optical domain, thereby reducing latency and energy consumption. To date, prism-coupled surface plasmon devices, two-dimensional or one-dimensional photonic crystals, and metasurfaces have been used for spatial differentiation, integration, and other optical analog computing [4–7].

Based on spatial-frequency filtering, metasurfaces have been widely used to perform optical analog operations, including first-order differentiation, second-order differentiation, and integration [8–11], and have been extended to applications such as edge detection, dual-polarization image processing, polarization imaging, quantum edge detection, and reconfigurable image processing [12-15]. In terms of polarization sensitivity, representative studies have mainly focused on polarization-independent or linear-polarization-selective image processing [16-22]. For example, Tanriover *et al*. experimentally demonstrated a polarization-independent all-optical edge-detection metasurface in the visible regime [16]. Cotrufo *et al*. realized polarization imaging and linear-polarization-selective directional edge detection [17]. However, these schemes offer limited degrees of freedom for polarization control, making flexible function switching and multichannel multiplexing difficult to achieve. To overcome this limitation, metasurfaces based on spiral-phase modulation or dynamic control have been used for dual-channel imaging. For example, Huo *et al*. realized spin-switchable bright-field imaging and spiral phase-contrast imaging [23]. Xiao *et al*. demonstrated electrically switchable bright-field and edge-enhanced imaging using a computational metasurface [24]. However, these schemes usually rely on local phase modulation and require a 4f system, which limits device compactness and integration capability. On the other hand, quasi-bound state in the continuum (quasi-BIC) or nonlocal metasurfaces can achieve high-quality edge detection [13, 15], but their high-Q resonances are usually accompanied by strong angular selectivity, thereby limiting the spatial-frequency modulation range and the effective numerical aperture.

To address these issues, we propose a transmissive terahertz image-processing scheme based on an extrinsically chiral all-dielectric metasurface. By exciting a spin-selective extrinsic chiral response under oblique incidence, the structure enables large-numerical-aperture spatial-frequency filtering at the same operating frequency, leading to edge-enhanced imaging under right-handed circularly polarized (RCP) incidence and bright-field imaging under left-handed circularly polarized (LCP) incidence, thus realizing dual-channel image processing.

Figure 1 shows the function and unit-cell geometry of the proposed all-dielectric metasurface with extrinsic chirality. As shown in Fig. 1(a), the metasurface operates in transmission mode and enables spin-selective image processing for RCP and LCP incidence when the terahertz wave impinges obliquely at a specific angle with respect to the surface normal. The RCP channel is used for edge detection, whereas the LCP channel preserves bright-field imaging. Figure 1(b) shows the structural parameters of the metasurface unit cell: the period is $a$ = 150 μm, and the top T-shaped silicon resonator has a thickness of $h_2$ = 150 μm with geometrical dimensions of $d_1$ = 145 μm, $d_2$ = 44 μm, $d_3$ = 70 μm, $d_4$ = 69 μm. The bottom layer is a $h_1$ = 300 μm-thick $SiO_2$ substrate, and the relative permittivities of Si and $SiO_2$ are 11.9 and 3.75, respectively.

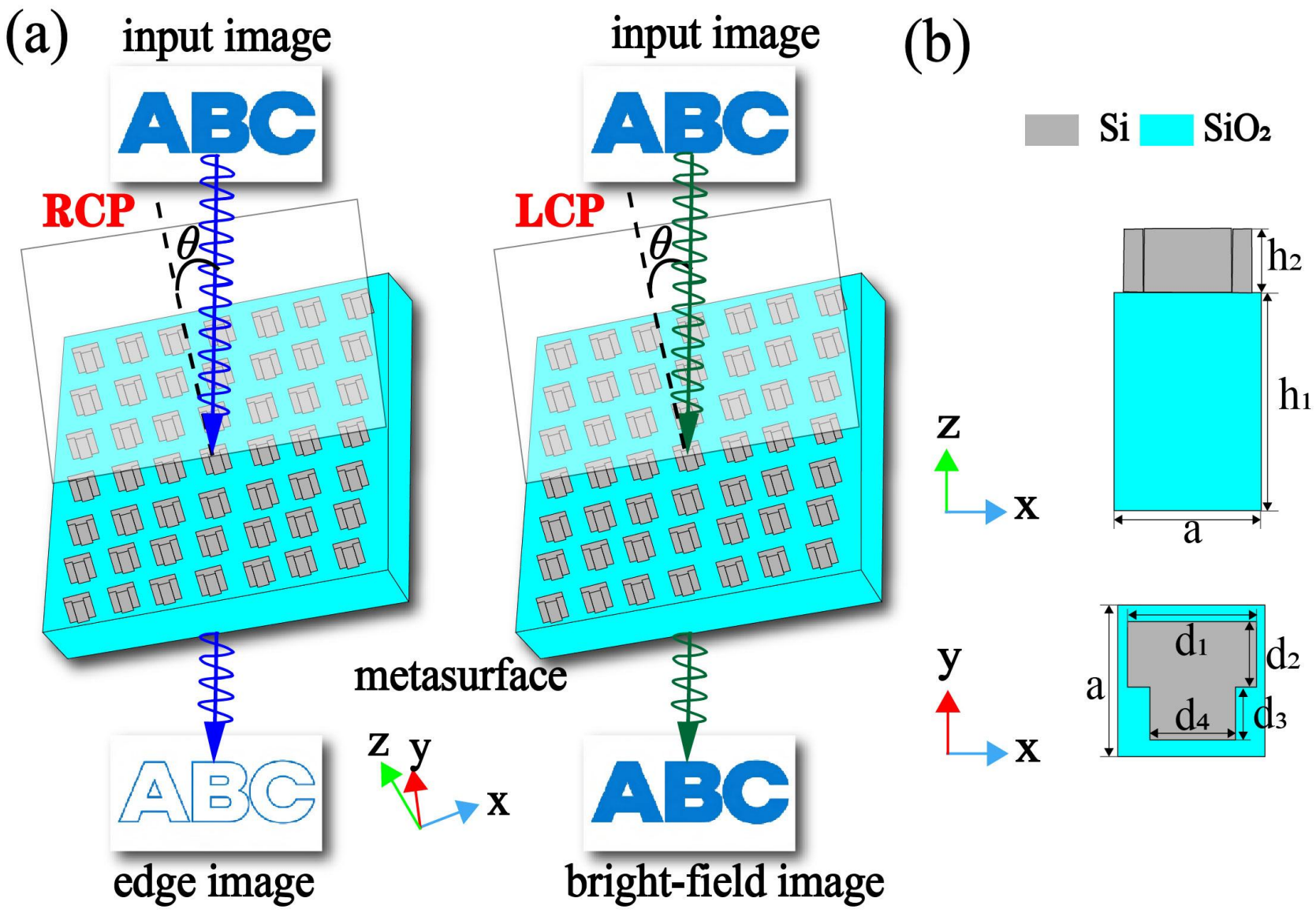

Fig. 1. Schematic of the function and unit-cell geometry of the all-dielectric metasurface with extrinsic chirality. (a) Circular-polarization-selective imaging under oblique incidence; (b) Structural parameters of the T-shaped metasurface unit cell.

First, we briefly analyze the design principle of the extrinsically chiral metasurface to reveal the physical mechanism underlying its circular-polarization-selective response. In the linear-polarization basis, the relation between the incident and transmitted electric fields can be described by a Jones matrix as follows:

$$\begin{bmatrix} E_x^t \\ E_y^t \end{bmatrix} = \begin{bmatrix} T_{xx} & T_{xy} \\ T_{yx} & T_{yy} \end{bmatrix} \begin{bmatrix} E_x^i \\ E_y^i \end{bmatrix} \tag{1}$$

Here, $E_x^i$ and $E_y^i$ are the incident electric-field components, $E_x^t$ and $E_y^t$ are the transmitted electric-field components, and $T_{ij}$ denotes the complex transmission coefficient from the incident polarization state $j$ to the transmitted polarization state $i$. By transforming this transmission matrix into the circular-polarization basis, it can be written as follows:

$$\begin{bmatrix} E_r^t \\ E_l^t \end{bmatrix} = \begin{bmatrix} T_{rr} & T_{rl} \\ T_{lr} & T_{ll} \end{bmatrix} \begin{bmatrix} E_r^i \\ E_l^i \end{bmatrix} \tag{2}$$

Here, $r$ and $l$ denote RCP and LCP waves, respectively. After the basis transformation, one obtains:

$$T_{rr,ll} = \frac{1}{2}\left(T_{xx} + T_{yy} \mp iT_{xy} \pm iT_{yx}\right) \tag{3}$$

It follows that the asymmetry of the cross-polarized transmission coefficients in the linear-polarization basis leads to different $T_{rr}$ and $T_{ll}$ in the circular-polarization basis, thereby giving rise to a circular-polarization-selective response.

To verify the above theoretical analysis, we further examine the extrinsic chiral response of the structure. Full-wave simulations were performed using commercial electromagnetic simulation software based on the finite-element method, where unit-cell boundary conditions were applied along the transverse *x* and *y* directions of the meta-atom, and open (add space) boundaries were used along the longitudinal *z* direction. Figures 2(a) and 2(b) show the circular-polarization transmission amplitudes under normal incidence ($\theta$= 0°) and oblique incidence ($\theta$= 30°), respectively. At normal incidence, the transmission responses for RCP and LCP waves differ only weakly, whereas $T_{rr}$ and $T_{ll}$ become clearly separated when the incidence angle increases to 30°, indicating that the extrinsic chiral response is effectively excited. The difference between $T_{rr}$ and $T_{ll}$ is most pronounced near $f_w$ = 0.69 THz, and this frequency is therefore selected as the operating frequency for the following analysis. To further reveal the physical origin of the circular-polarization selectivity, Figures. 2(c) and 2(d) compare the multipole scattering powers under oblique RCP and LCP incidence, showing that the magnetic dipole (MD) contribution dominates near the operating frequency. Compared with LCP incidence, the MD scattering power is significantly enhanced under RCP incidence, indicating stronger modal coupling of the unit cell to RCP waves. Figures 2(e) and 2(f) show the magnetic-field and displacement-current distributions in the $z = h_2/2$ plane, plotted using the same color scale, where the white arrows indicate the direction of the displacement-current density. Under RCP incidence, the field energy is mainly localized at the transverse arm and edge regions of the T-shaped resonator, and the displacement current exhibits pronounced local counter-rotating circulation, indicating that a strong MD resonance is efficiently excited. In contrast, under LCP incidence, both the field localization and circulating-current features are markedly weakened, indicating weaker coupling to the MD mode. This result agrees well with the multipole scattering analysis and confirms that the MD mode plays an important role in the circular-polarization-selective response under oblique incidence.

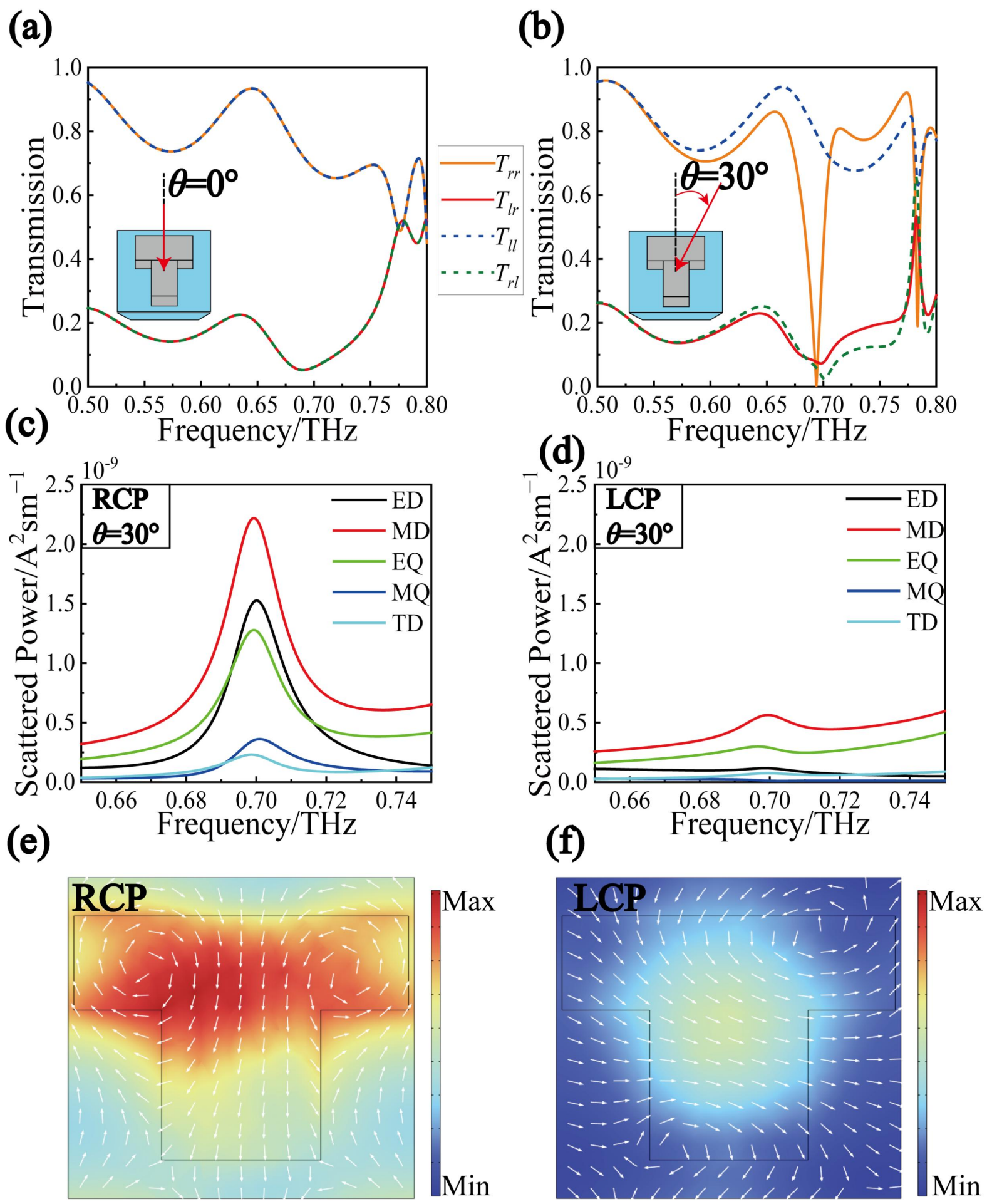

Fig. 2. Extrinsic chiral response of the metasurface. (a),(b) Circular-polarization transmission amplitudes under normal and oblique incidence; (c),(d) Multipole scattering powers under oblique RCP and LCP incidence; (e),(f) Magnetic-field and displacement-current distributions in the $z=h_2/2$ plane under RCP and LCP incidence.

Based on the above spin-selective response, we next analyze the edge-detection mechanism of the metasurface from the perspective of spatial-frequency filtering. Image edges usually correspond to regions where the grayscale or intensity varies rapidly in space and are mainly associated with high-spatial-frequency components. In contrast, flat regions with slowly varying grayscale or intensity mainly contain low-spatial-frequency components. Therefore, edge detection can essentially be regarded as a spatial high-pass filtering process. Let the input optical field be $E_{in}(x, y)$, whose spatial spectrum after Fourier transformation is $E_{in}(k_x, k_y)$. After passing through an optical system with spatial-filtering functionality, the output spectrum is given by $E_{out}(k_x, k_y)=H(k_x, k_y)E_{in}(k_x, k_y)$, where $H(k_x, k_y)$ is the spatial transfer function, $k_x$ and $k_y$ are the transverse spatial wave-vector components, and $k_x a/2\pi$ and $k_y a/2\pi$ are used for normalization. Therefore, when $H(k_x, k_y)$ suppresses low spatial frequencies while transmitting high spatial frequencies, the edge information in the output image is enhanced.

Figure 3 shows the spatial-filtering characteristics of the metasurface under RCP incidence. Figure 3(a) presents the transmission spectra as functions of incidence angle ($\theta$) and frequency ($f$), where the dashed line marks the operating frequency ($f_w$). At the operating frequency, the transmission amplitude undergoes a pronounced transition from nearly complete transmission to strong suppression as the incidence angle increases to $\theta$=30°.

Figures 3(b) and 3(c) show the transmission-amplitude and phase distributions at the operating frequency for incidence angles from 0° to 60°. This angular range is centered at $\theta$=30°, corresponding to a maximum effective numerical aperture of $NA_{max}$=sin (60°−30°)=0.5, indicating that the metasurface can achieve angle-dependent transmission control over a large spatial-frequency range. Near $k_{CR}$ = +0.1736$k_0$, where $k_{CR}$ denotes the central transverse wave-vector shift induced by oblique incidence, the transmission amplitude rapidly decreases from nearly 1 to nearly 0, accompanied by a finite phase variation, indicating pronounced spatial-frequency selectivity in this region. Since $H(k_x, k_y)$ is determined by both the transmission amplitude and phase, the amplitude suppression and finite phase variation near this region indicate that the metasurface can effectively modulate spatial-frequency components, providing the basis for constructing the optical transfer function (OTF). Furthermore, to quantitatively evaluate the chiral selectivity and performance of the metasurface, the circular dichroism is calculated and defined as follows:

$$CD = \left(\left|T_{rr}\right|^2 + \left|T_{lr}\right|^2\right) - \left(\left|T_{rl}\right|^2 + \left|T_{ll}\right|^2\right) \tag{4}$$

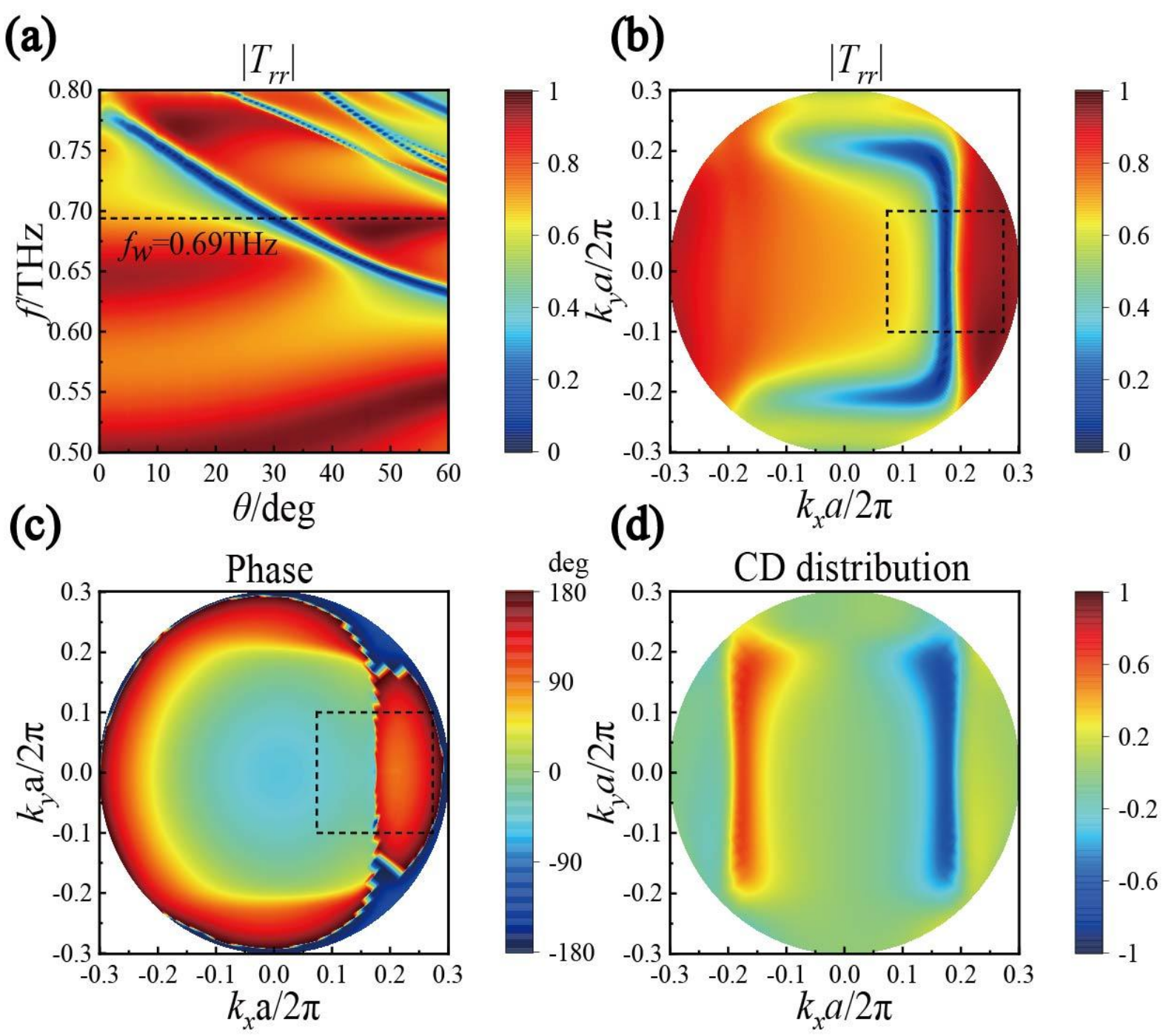

Figure 3. Spatial filtering characteristics of the metasurface: (a) transmission spectra as functions of incidence angle ($\theta$) and frequency ($f$); (b)–(d) transmission amplitude, transmission phase, and CD distributions at the operating frequency ($f_w$).

As shown in Fig. 3(d), the peak $CD$ reaches approximately 0.7 at the operating frequency, indicating a pronounced extrinsic chiral response of the structure. Moreover, owing to the opposite handedness of the two circular polarization states, the $CD$ exhibits a sign reversal in momentum space, confirming the distinct transmission responses under RCP and LCP incidence.

Based on the above analysis, the optical transfer function at the operating frequency is further extracted. Since the transmission-amplitude suppression and phase variation under RCP incidence mainly occur near $\theta$= 30°, the corresponding momentum coordinate $k_{CR}$ is used as the reference point, and the local momentum coordinates are written as ($k_xa/2\pi-k_{CR}$, $k_ya/2\pi$). The OTF is restricted to a finite field-of-view region satisfying $|(k_xa/2\pi-k_{CR})| \leq 0.1$ and $|k_ya/2\pi| \leq 0.1$, as indicated by the dashed boxes in Figs. 3(b) and 3(c), corresponding to a maximum numerical aperture of $NA_{OTF,\ max}\approx 0.4$, which reflects the spatial-frequency filtering capability of the metasurface under large-NA conditions. Figures 4(a), 4(b), 4(c), and 4(d) show the transmission-amplitude and phase distributions within this field of view under RCP and LCP incidence, respectively. The OTF profiles along the horizontal and vertical directions are further extracted, as shown in Figs. 4(e) and 4(f), indicating that under RCP incidence, the horizontal OTF exhibits a pronounced Lorentzian-like profile, which suppresses the central low-spatial-frequency components while transmitting higher-spatial-frequency components, thereby enabling edge detection. In contrast, under LCP incidence, the corresponding OTF profiles allow most spatial-frequency components to pass through the metasurface, thus supporting bright-field imaging. These results demonstrate that the metasurface can switch between edge detection and bright-field imaging within the same structure, at the same operating frequency and incidence angle, by changing the incident circular polarization state, thereby improving the device integration and modulation flexibility.

To verify the practical image-processing capability of the designed metasurface, Figures. 5(a) and 5(b) show the input images, each consisting of four 300×300-pixel images. And under RCP incidence; (c) horizontal and vertical OTF amplitude profiles under RCP incidence; (d)–(f) the corresponding results for LCP incidence.

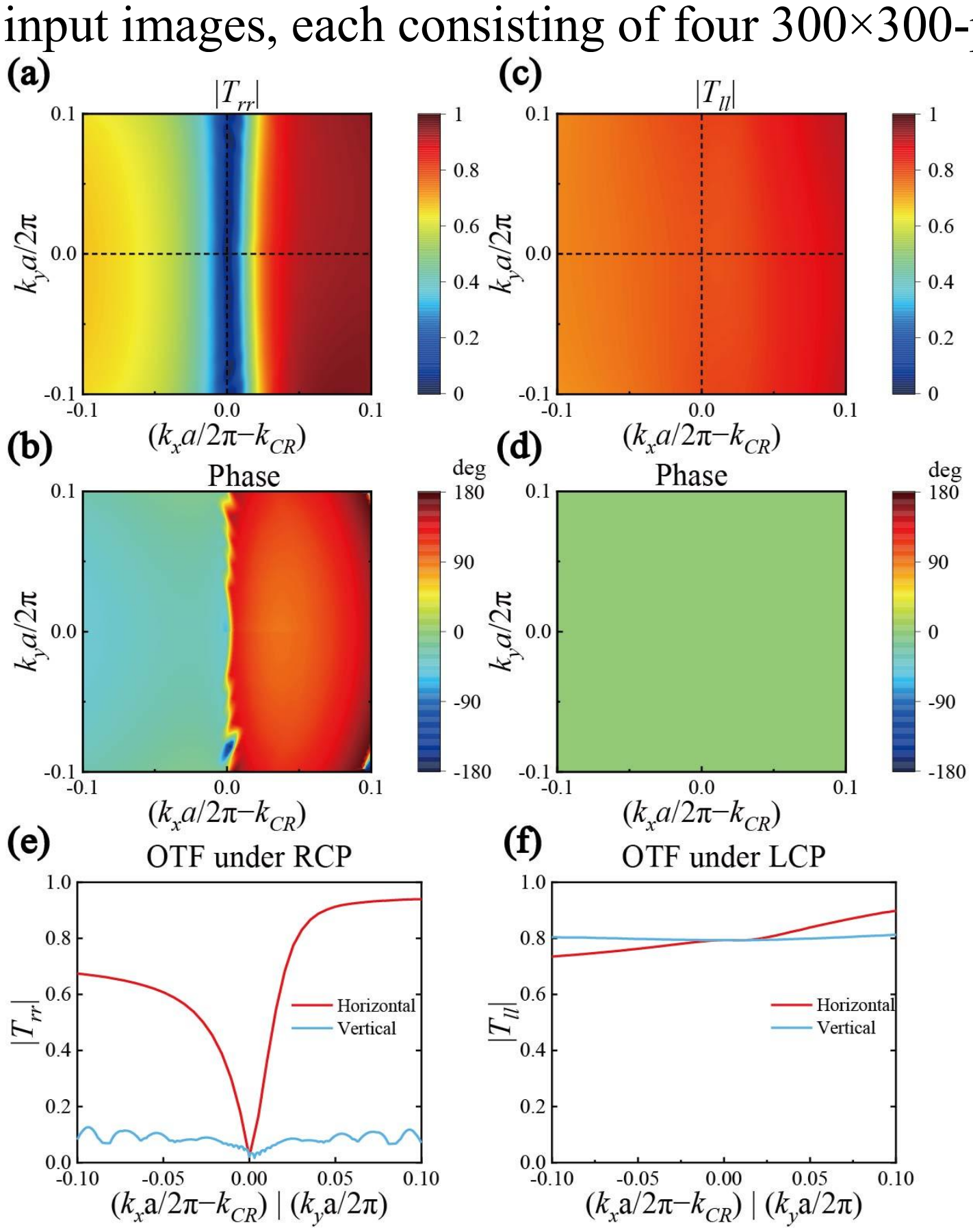

Figure 4. Filtering functions at the operating frequency $f_w$=0.69 THz within the field-of-view region satisfying $|(k_xa/2\pi-k_{CR})| \leq 0.1$ and $k_ya/2\pi| \leq 0.1$: (a), (b) transmission amplitude and phase

Figures 5(c) and 5(d) present the output results at the operating frequency under an incidence angle of $\theta = 30°$, showing pronounced edge enhancement under RCP incidence, whereas the output image retains bright-field imaging features under LCP incidence. To further quantitatively compare the image changes before and after filtering, normalized intensity distributions are extracted along the central horizontal line of each output image, as shown in Figures. 5(e) and 5(f). Under RCP incidence, the filtered intensity peaks mainly appear at the pattern boundaries, whereas the intensity is nearly zero in regions away from the edges, indicating that the low-spatial-frequency background information is suppressed while the high-spatial-frequency edge information is enhanced. In contrast, under LCP incidence, the intensity profiles before and after filtering are nearly identical, indicating that the main image information is preserved, corresponding to bright-field imaging. These image-simulation results further verify that the metasurface can switch its functionality by changing the incident circular polarization state.

In summary, this work demonstrates transmission-type terahertz spin-selective image processing based on an extrinsic-chiral all-dielectric metasurface. Through spatial coupling between the oblique incident wave vector and the T-shaped dielectric meta-atom, the structure produces a spin-dependent transmission response near 0.69 THz and provides effective spatial-frequency modulation over a broad angular-spectrum range. The resulting spin-dependent filtering functions enable edge-enhanced imaging under RCP incidence while preserving bright-field imaging under LCP incidence. These results indicate that our work provides a new route for designing transmission-type terahertz all-optical image-processing devices with circular-polarization selectivity and large-numerical-aperture characteristics.

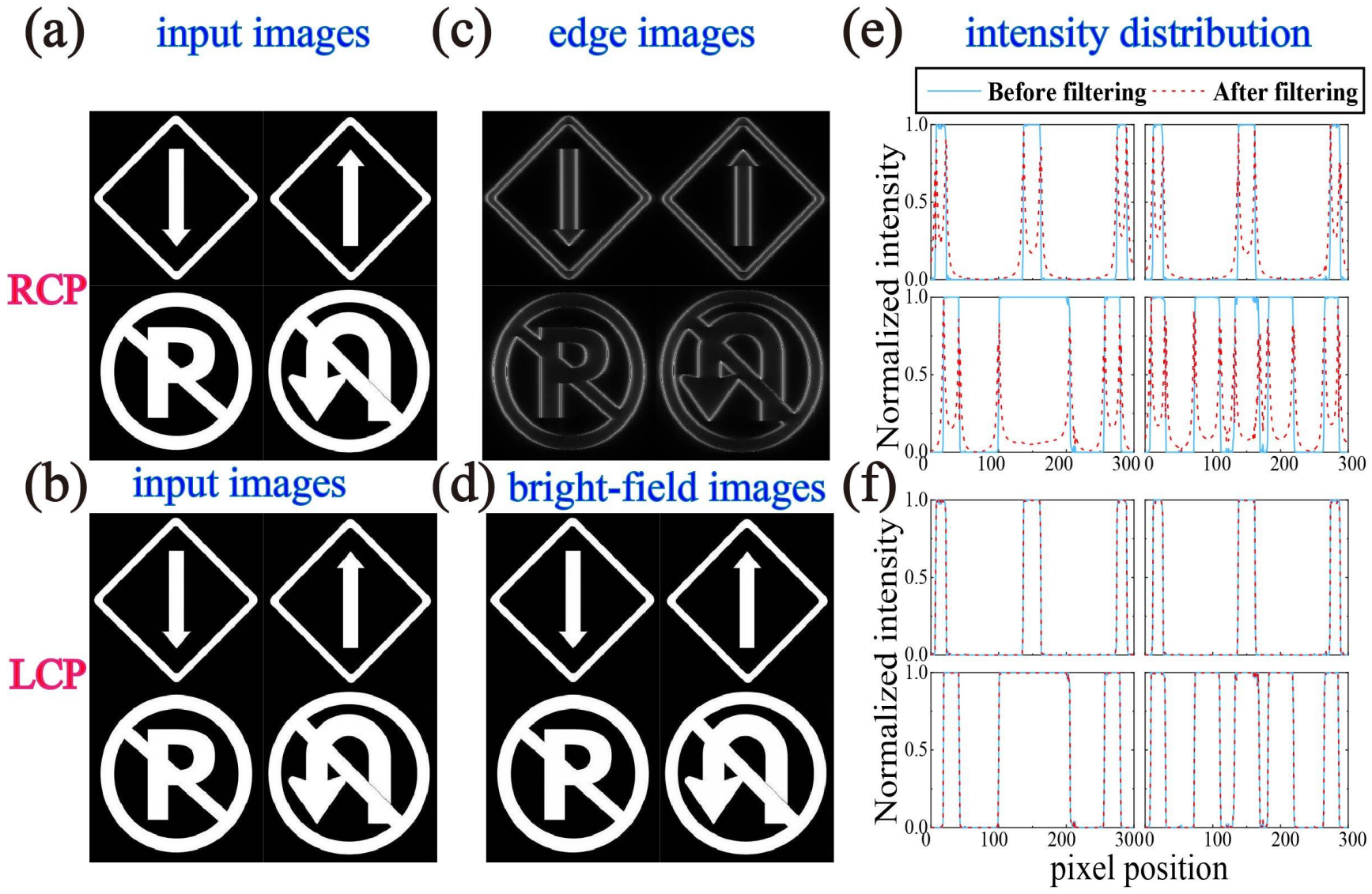

Figure 5. Image-processing results: (a), (b) input images; (b), (e) edge-enhanced images under RCP incidence and bright-field images under LCP incidence, respectively; (c), (f) normalized intensity distributions extracted along the central horizontal lines of the corresponding output images.